\documentclass[sn-nature]{sn-jnl}

\usepackage{graphicx}%
\usepackage{multirow}%
\usepackage{amsmath,amssymb,amsfonts}%
\usepackage{amsthm}%
\usepackage{mathrsfs}%
\usepackage[title]{appendix}%
\usepackage{xcolor}%
\usepackage{textcomp}%
\usepackage{manyfoot}%
\usepackage{booktabs}%
\usepackage{algorithm}%
\usepackage{algorithmicx}%
\usepackage{algpseudocode}%
\usepackage{listings}%
\usepackage{placeins}
\usepackage{url}

\newcommand{\Jzero}{ZTF\,J0440+2325}
\newcommand{\Jone}{ZTF\,J1444+4820}

\theoremstyle{thmstyleone}%

\theoremstyle{thmstyletwo}%

\theoremstyle{thmstylethree}%

\begin{document}

\title[Stars stably accreting from substellar objects]{Stars stably accreting from substellar objects}

%%=============================================================%%
%% Authors and affiliations
%%=============================================================%%

\author*[1,2]{\fnm{Aaron} \sur{Householder}}
\email{aaron593@mit.edu}

\author[3,2,4]{\fnm{Kaitlyn} \sur{Shin}}

\author[3,2]{\fnm{Kevin B.} \sur{Burdge}}

\author[]{\fnm{Thomas R.} \sur{Marsh}}

\author[3,2]{\fnm{Saul A.} \sur{Rappaport}}

\author[4]{\fnm{Kareem} \sur{El-Badry}}

\author[3,2]{\fnm{Joheen} \sur{Chakraborty}}

\author[3,2]{\fnm{Emma} \sur{Chickles}}

\author[5]{\fnm{Fei} \sur{Dai}}

\author[4]{\fnm{Matthew J.} \sur{Graham}}

\author[4]{\fnm{S.R.} \sur{Kulkarni}}

\author[6,7]{\fnm{Pablo} \sur{Rodríguez-Gil}}

\author[8]{\fnm{Andrew} \sur{Vanderburg}}

\author[4]{\fnm{Samuel} \sur{Whitebook}}

\affil[1]{%
\orgdiv{Department of Earth, Atmospheric and Planetary Sciences},
\orgname{Massachusetts Institute of Technology},
\orgaddress{\city{Cambridge}, \state{MA}, \postcode{02139}, \country{USA}}
}

\affil[2]{%
\orgdiv{Kavli Institute for Astrophysics and Space Research},
\orgname{Massachusetts Institute of Technology},
\orgaddress{\city{Cambridge}, \state{MA}, \postcode{02139}, \country{USA}}
}

\affil[3]{%
\orgname{Department of Physics},
\orgname{Massachusetts Institute of Technology},
\orgaddress{\city{Cambridge}, \state{MA}, \postcode{02139}, \country{USA}}
}

\affil[4]{%
\orgdiv{Division of Physics, Mathematics, and Astronomy},
\orgname{California Institute of Technology},
\orgaddress{\city{Pasadena}, \state{CA}, \postcode{91125}, \country{USA}}
}

\affil[5]{%
\orgname{Institute for Astronomy, University of Hawai`i},
\orgaddress{\street{2680 Woodlawn Drive}, \city{Honolulu}, \state{HI}, \postcode{96822}, \country{USA}}
}

\affil[6]{%
\orgname{Instituto de Astrofísica de Canarias},
\orgaddress{\postcode{E-38205}, \city{La Laguna}, \state{Tenerife}, \country{Spain}}
}

\affil[7]{%
\orgdiv{Departamento de Astrofísica},
\orgname{Universidad de La Laguna},
\orgaddress{\postcode{E-38206}, \city{La Laguna}, \state{Tenerife}, \country{Spain}}
}

\affil[8]{%
\orgname{Center for Astrophysics | Harvard \& Smithsonian},
\orgaddress{\street{60 Garden Street}, \city{Cambridge}, \state{MA}, \postcode{02138}, \country{USA}}
}

%%=============================================================%%
%% Abstract from original Nature submission
%%=============================================================%%

\abstract{Substellar objects such as brown dwarfs and planets are generally expected to remain detached from their main-sequence host stars unless orbital decay or stellar expansion brings them into contact, leading to rapid engulfment and destruction (e.g., \cite{De2023}). Such a fate is predicted for the Earth and other rocky planets in our solar system \cite{Schroder2008}; however, in certain cases, theory also allows for stable long-lived mass transfer from a substellar object onto its main-sequence host \cite{Hjellming1987}, though such accretion has never been directly observed. Here we report the first direct observations of stable mass transfer from a substellar object onto a main-sequence star. In particular, we identify two binaries, \Jzero\ and \Jone, with orbital periods of just 87 and 67 minutes, respectively, in which a brown dwarf stably transfers mass onto an M dwarf companion. These systems demonstrate that the fate of some substellar objects is not rapid engulfment and destruction, but instead gradual consumption for potentially billions of years.}
\maketitle

%%=============================================================%%
%% Main text from updated Science format
%%=============================================================%%

\begin{figure}
    \centering
    \includegraphics[width=0.9\textwidth]{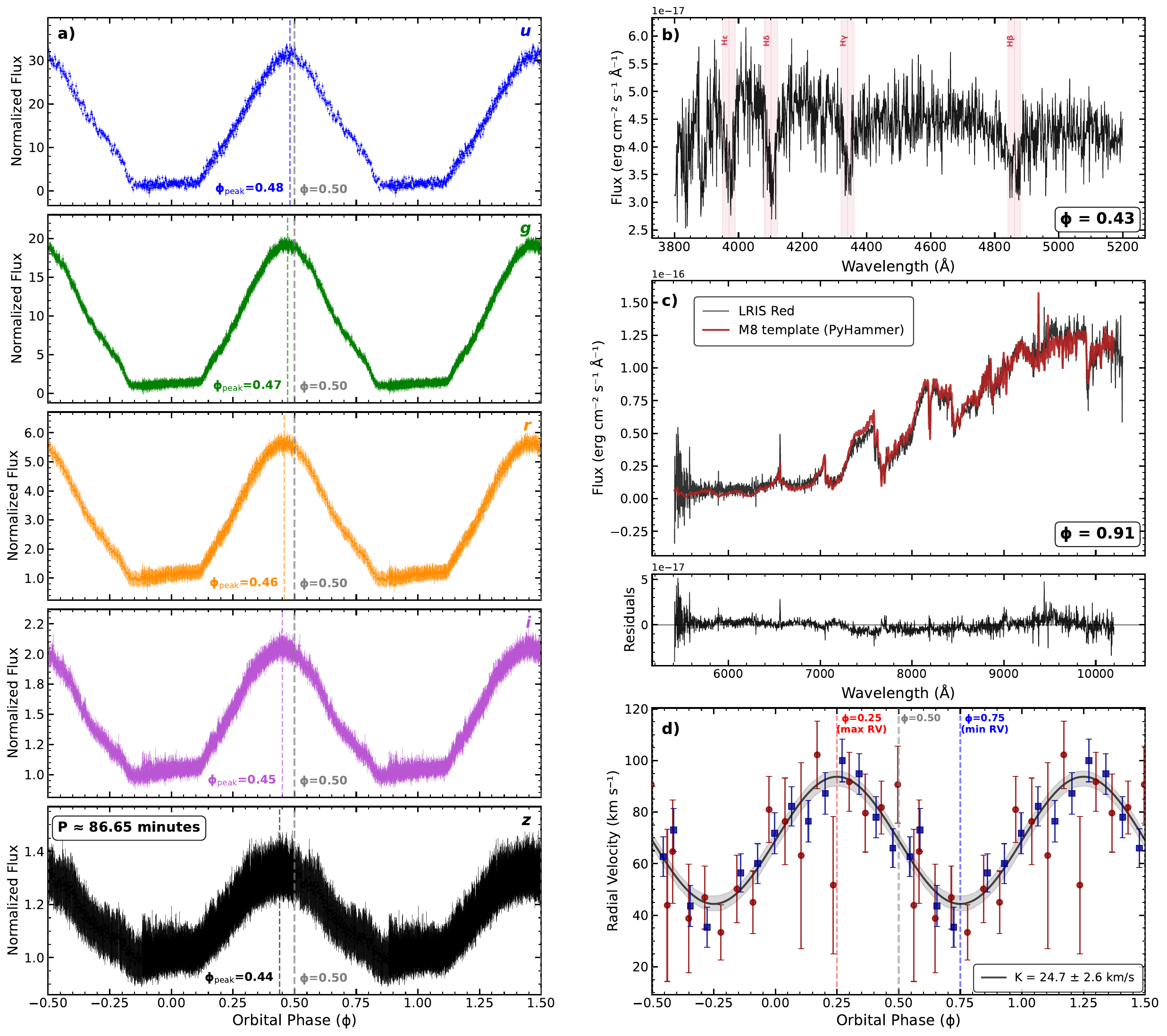}
    \caption{\textbf{Photometric and spectroscopic observations of \Jzero.}
    (\textbf{a}) Phase-folded HiPERCAM light curves in the $u$, $g$, $r$, $i$, and $z$ bands, showing large-amplitude chromatic variability produced by an accretion hot spot on the M dwarf.
    (\textbf{b}) Keck/LRIS blue-arm spectrum obtained near peak brightness ($\phi = 0.43$), with transient Balmer absorption arising from the accretion hot spot.
    (\textbf{c}) Keck/LRIS red-arm spectrum obtained near minimum brightness ($\phi = 0.91$), showing molecular absorption features consistent with an M8 dwarf template (red).
    (\textbf{d}) Radial-velocity measurements of the M dwarf from the ESI spectra (blue) and LRIS spectra (red) as a function of orbital phase. The best-fitting radial velocity model has a semi-amplitude of $K = 24.7 \pm 2.6\,\mathrm{km\,s^{-1}}$, demonstrating that the unseen companion lies in the substellar mass regime.}
    \label{fig:J0440}
\end{figure}

\noindent
We identified \Jzero\ in a systematic search for short-period variables in data from the Zwicky Transient Facility (ZTF) \cite{Bellm2019,Graham2019,Masci2019}. The system had a highly significant periodic signal at an orbital period of $P = 86.65$ minutes, which motivated ground-based follow-up. On UT 2023 October 18, we obtained high-speed simultaneous five-band ($u,g,r,i,z$) photometry of \Jzero\ with HiPERCAM \cite{Dhillon2021} on the $10.4$-m Gran Telescopio Canarias (GTC).  The light curves showed extreme chromatic variability, with amplitudes that increased toward shorter wavelengths (Figure~\ref{fig:J0440}). In particular, \Jzero\ brightened by more than a factor of $30$ in the $u$ band relative to minimum brightness, while the corresponding increase in brightness in the $z$ band was comparatively modest (a factor of roughly $1.4$). Such short period and large wavelength-dependent modulations are not produced by isolated late-type stars, but are characteristic of compact binaries in which one component is actively accreting material from a companion and develops a bright hot spot that rotates in and out of view \cite{Warner1995}. Indeed, light-curve modeling with \texttt{lcurve} \cite{Copperwheat2010} showed that the HiPERCAM data are well described by a large self-eclipsing hot spot on one component of the binary (Methods). However, the photometry alone does not reveal the nature of the binary or what physical process is powering the hot spot.

To better understand the photometric behavior of \Jzero, we obtained phase-resolved spectroscopy with the Low Resolution Imaging Spectrometer (LRIS) on the $10$-m Keck I telescope \cite{Oke1995} as well as the Echellette Spectrograph and Imager (ESI) on the Keck II telescope \cite{Sheinis2002}. The spectra of \Jzero\ had M dwarf absorption features and showed excellent agreement with the empirical M8 dwarf template from \texttt{PyHammer} \cite{Kesseli2017} (Figure~\ref{fig:J0440}). We also measured the radial velocities of the M dwarf and found a semi-amplitude of $K = 24.7 \pm 2.6\,\mathrm{km\,s^{-1}}$ (Methods), placing the unseen companion in the substellar mass regime.  For instance, adopting a typical late-M dwarf mass of $M=0.11\,M_\odot$ and numerically solving the exact mass-function relation yields a minimum companion mass of $M\sin i=11.7\pm1.3\,M_{\mathrm{Jup}}$. Combined with the strong optical variability, we therefore conclude that \Jzero\ is a mass-transferring M dwarf--brown-dwarf binary.

\begin{figure}
    \centering
    \includegraphics[height=0.85\textheight]{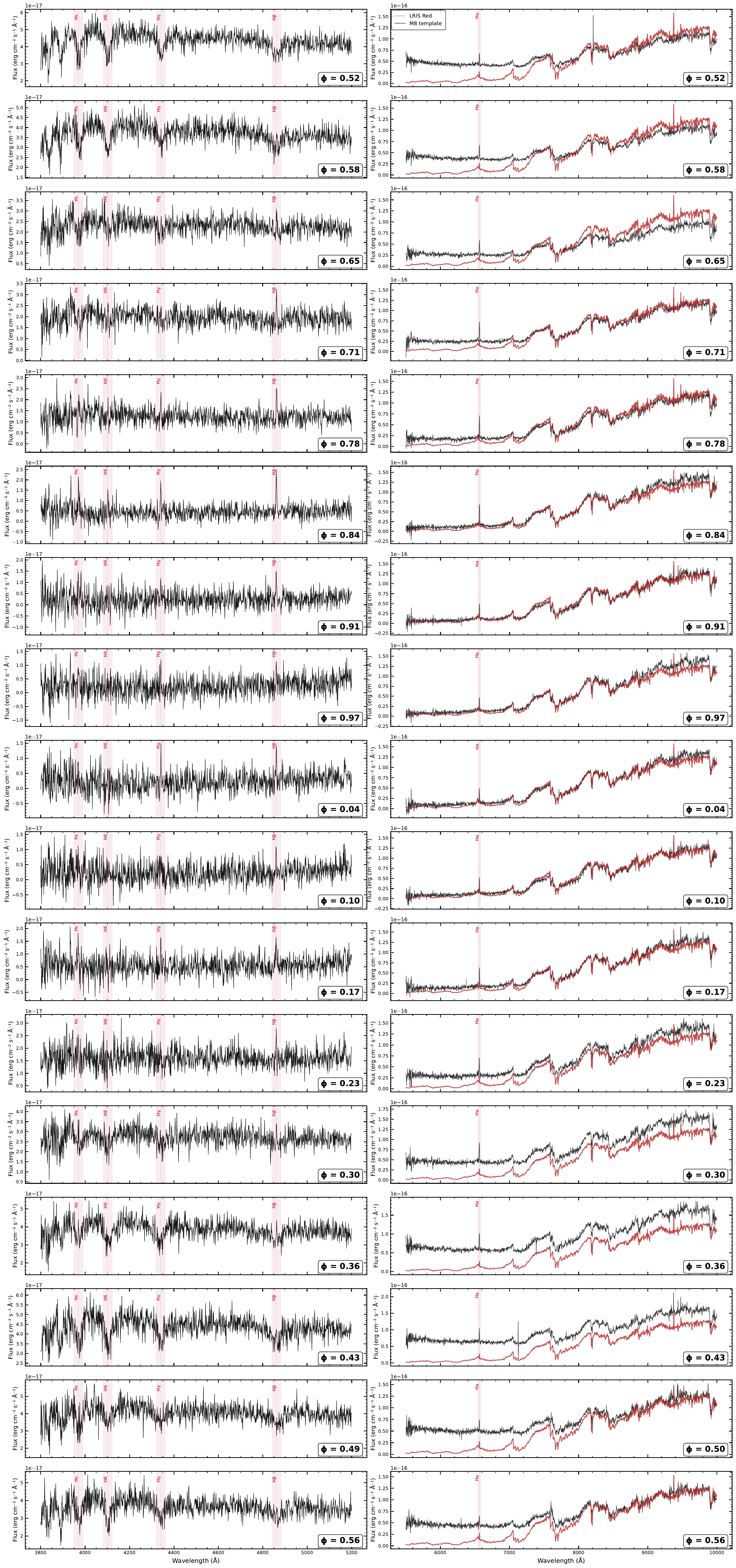}
    \caption{\textbf{Full phase-resolved Keck/LRIS spectra of \Jzero.}
    The red-arm spectra show molecular absorption consistent with an M8 dwarf template (red), while transient Balmer absorption appears near peak brightness in the blue arm spectra. The empirical M8 dwarf template from \texttt{PyHammer} is scaled to match the red-arm spectrum obtained near minimum optical brightness ($\phi = 0.91$) and is overplotted at all orbital phases with the same normalization.}
    \label{fig:J0440keck}
\end{figure}

This interpretation is also supported by the phase-resolved LRIS blue-arm spectra of \Jzero, which are largely featureless apart from transient Balmer absorption near peak optical brightness (Extended Data Figure~\ref{fig:J0440keck}). Such behavior is consistent with a buried accretion shock, in which the stream penetrates beneath the photosphere of the accretor to the depth where its ram pressure matches the atmospheric pressure of the accretor \cite{Marsh2002}. The resulting heated photosphere produces transient Balmer-absorption--dominated spectra, analogous to those observed on the dayside of many strongly heated companions in black widow pulsars (e.g., \cite{Romani2021}).

One remarkable aspect of \Jzero\ is that both the phase-dependent Balmer absorption and the overall light-curve morphology are extremely similar to black widow pulsar systems, where a neutron star irradiates a low-mass companion.  Indeed, \cite{Burdge2022} identified ZTF J1406+1222, a $62$-minute binary with very similar optical variability, transient Balmer features, and a small radial velocity semi-amplitude and interpreted it as a black widow pulsar. However, this black widow pulsar scenario is strongly disfavored for \Jzero\ for multiple reasons. Most importantly, the measured radial-velocity semi-amplitude of the M dwarf is $K = 24.7 \pm 2.6\,\mathrm{km\,s^{-1}}$, which would require an orbital inclination $i \le 2.5^\circ$ if the unseen companion were a $1.4\,M_\odot$ neutron star, corresponding to an a priori probability of only $9\times10^{-4}$ (Methods). Furthermore, \Jzero\ is not detected in X-rays, with a $3\sigma$ X-ray luminosity upper limit of $ 7.1\times10^{29}\,\mathrm{erg\,s^{-1}}$ (Methods), below the luminosities typically observed from black widow pulsars ($L_X\sim10^{30}$--$10^{32}\,\mathrm{erg\,s^{-1}}$) \cite{Gentile2014}.

\begin{figure}
    \centering
    \includegraphics[width=0.9\textwidth]{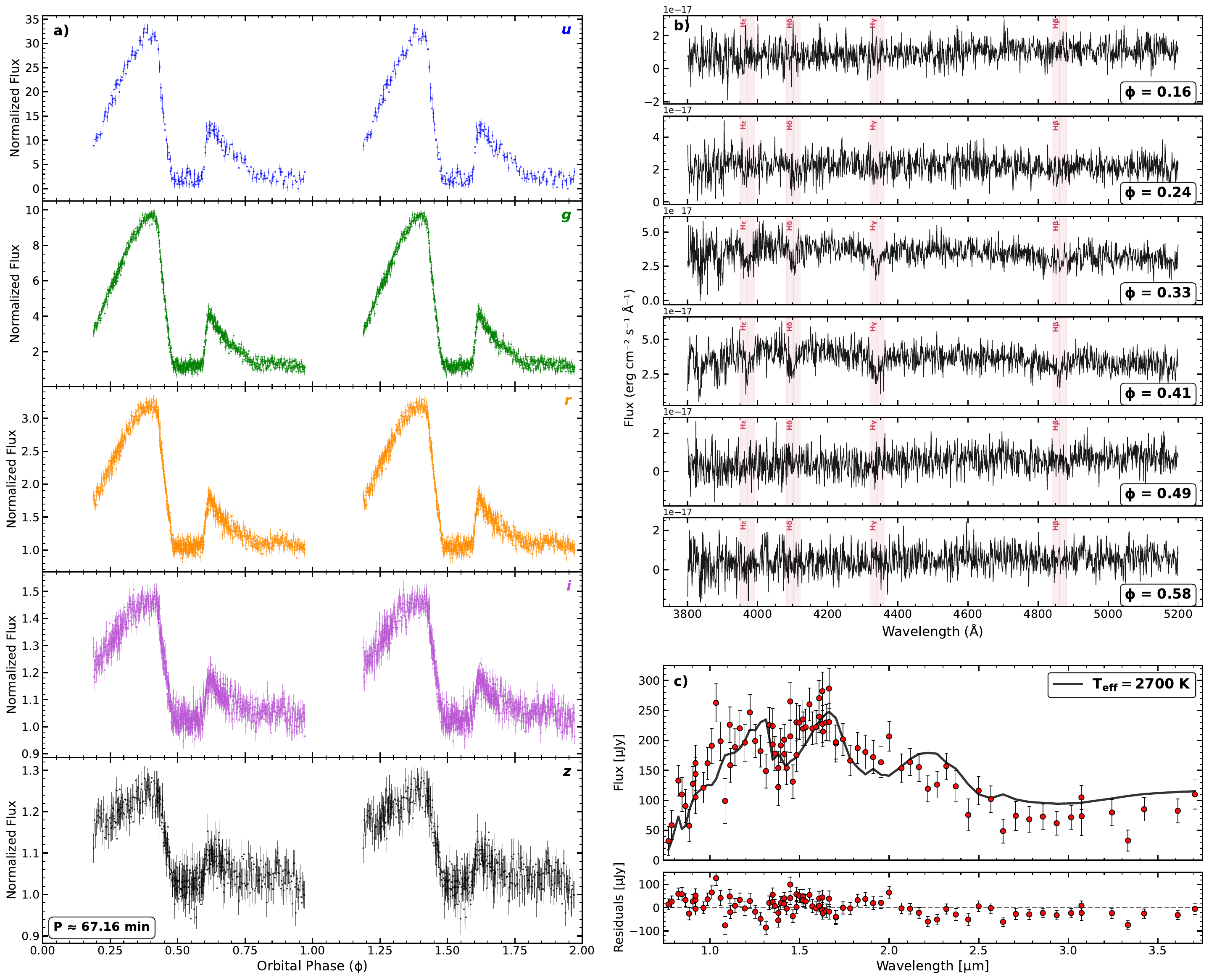}
    \caption{\textbf{Photometric and spectroscopic observations of \Jone.}
    (\textbf{a}) Phase-folded HiPERCAM light curves of \Jone\ in the $u$, $g$, $r$, $i$, and $z$ bands, showing strong wavelength-dependent variability from an accretion hot spot on the M dwarf that is eclipsed by the brown dwarf companion near peak brightness.
    (\textbf{b}) Phase-resolved Keck/LRIS blue-arm spectra obtained over a portion of the orbit, with transient Balmer absorption appearing near peak brightness when the hot spot dominates the optical flux.
    (\textbf{c}) \textit{SPHEREx} near-infrared spectro-photometry covering $0.75$--$3.75\,\mu\mathrm{m}$ (red points with error bars) compared to a BT--Settl atmosphere model with $T_{\mathrm{eff}} = 2700\,\mathrm{K}$ (black line), indicating that the dominant photospheric component is indeed a late-type M dwarf.
    }
    \label{fig:J1444}
\end{figure}

In addition to \Jzero, we identified \Jone, a second brown dwarf--M dwarf mass-transferring system with nearly analogous photometric behavior that definitively cannot be a black widow pulsar. \Jone\ has an orbital period of $P = 67.16$ minutes and showed eclipses in its ZTF light curves. On UT 2024 August 3, we obtained high-speed simultaneous five-band  photometry of \Jone\ with HiPERCAM, which confirmed the eclipse and the same extreme wavelength-dependent variability seen in \Jzero, with very large amplitudes in the blue and much smaller amplitudes in the red (Figure~\ref{fig:J1444}). As in \Jzero, the HiPERCAM light curves of \Jone\ are also well fit by a hot spot (Methods), with the primary difference being that the hot spot in \Jone\ is eclipsed by the companion near peak brightness (Extended Data Figure~\ref{fig:J1444lcurve}). In black widow pulsar systems, the optical variability is entirely driven by irradiation of the companion, so such systems do not produce sharp eclipses in the optical.

While the presence of eclipses strongly rules out a black widow pulsar interpretation for \Jone, the eclipses do not impose the usual geometric and dynamical constraints associated with eclipsing binaries. This is because the eclipses correspond to occultations of the hot spot, not of the underlying stellar photospheres. To further characterize the system, we thus obtained phase-resolved Keck/LRIS spectroscopy of \Jone. However, a focusing issue prevented us from obtaining usable red-arm spectra, and the available blue-arm data show only transient Balmer absorption near peak brightness (which help constrain the nature of the system but do not provide stable photospheric features for phase-resolved radial-velocity measurements). As a result, the component masses and spectral types of \Jone\ cannot yet be dynamically constrained as directly as in \Jzero.

Nonetheless, the \textit{SPHEREx} spectrum of \Jone\ shows clear evidence for a late-type photosphere consistent with an M dwarf, suggesting that one component in the binary is indeed an M dwarf (Figure~\ref{fig:J1444}; Extended Data Figure~\ref{fig:spherex}). While we do not yet have radial velocities, the ultra-short orbital period of \Jone\ severely restricts the nature of the unseen companion, as only objects with very high mean densities can exist at such small separations (see e.g., \cite{Rappaport2021}). In particular, at this orbital period the unseen companion to the M-dwarf cannot be a hydrogen-burning star \cite{Rappaport1983,Rappaport2021}, so the remaining possible companions are limited to highly dense objects such as brown dwarfs or compact objects (e.g., white dwarfs or neutron stars). The compact-object interpretations are ruled out by the lack of X-ray emission ($L_X < 7.1\times10^{29}\,\mathrm{erg\,s^{-1}}$), the optical spectra showing only transient Balmer features, and the presence of eclipses (Methods). We therefore conclude that \Jone\ is the same class of object as \Jzero: a brown-dwarf--M dwarf binary undergoing active mass transfer.

While the evidence above indicates that \Jone\ and \Jzero\ are mass-transferring brown-dwarf--M dwarf binaries, it does not immediately distinguish which component is the donor and which is the accretor. However, for low-mass convective donors, classical theory predicts that stable Roche-lobe overflow requires mass ratios $q = M_{\rm don}/M_{\rm acc} \lesssim 2/3$ \cite{Rappaport1983,Hjellming1987}, indicating that a configuration in which the brown dwarf is the accretor would be dynamically unstable (Methods). Assuming that the M dwarf is indeed the accretor in both systems, we can use the Roche geometry to constrain the component masses and radii to test this stability criterion (Methods). At their measured orbital periods, both \Jzero\ and \Jone\ comfortably satisfy the requirement for stable mass transfer, with posterior mass ratios $q = 0.27^{+0.06}_{-0.04}$ and $q = 0.41^{+0.07}_{-0.06}$, respectively (Table~\ref{table1}). Furthermore, the discovery of these two systems also provides an empirical argument that the mass-transfer is stable and not short-lived. Very few short-period detached M dwarf--brown dwarf binaries are known, and reaching such ultrashort periods is expected to require several Gyr of orbital evolution \cite{ElBadry2023}. Thus, if the accretion led to rapid instability, the likelihood of observing two systems in this state would be extremely small. We therefore conclude that mass transfer in these systems proceeds from the brown dwarf to the M dwarf and is long-lived, likely lasting at least $\sim10^{7}$--$10^{8}$ yr and plausibly hundreds of Myr to several Gyr.

\begin{figure}
    \centering
    \includegraphics[width=0.6\textwidth]{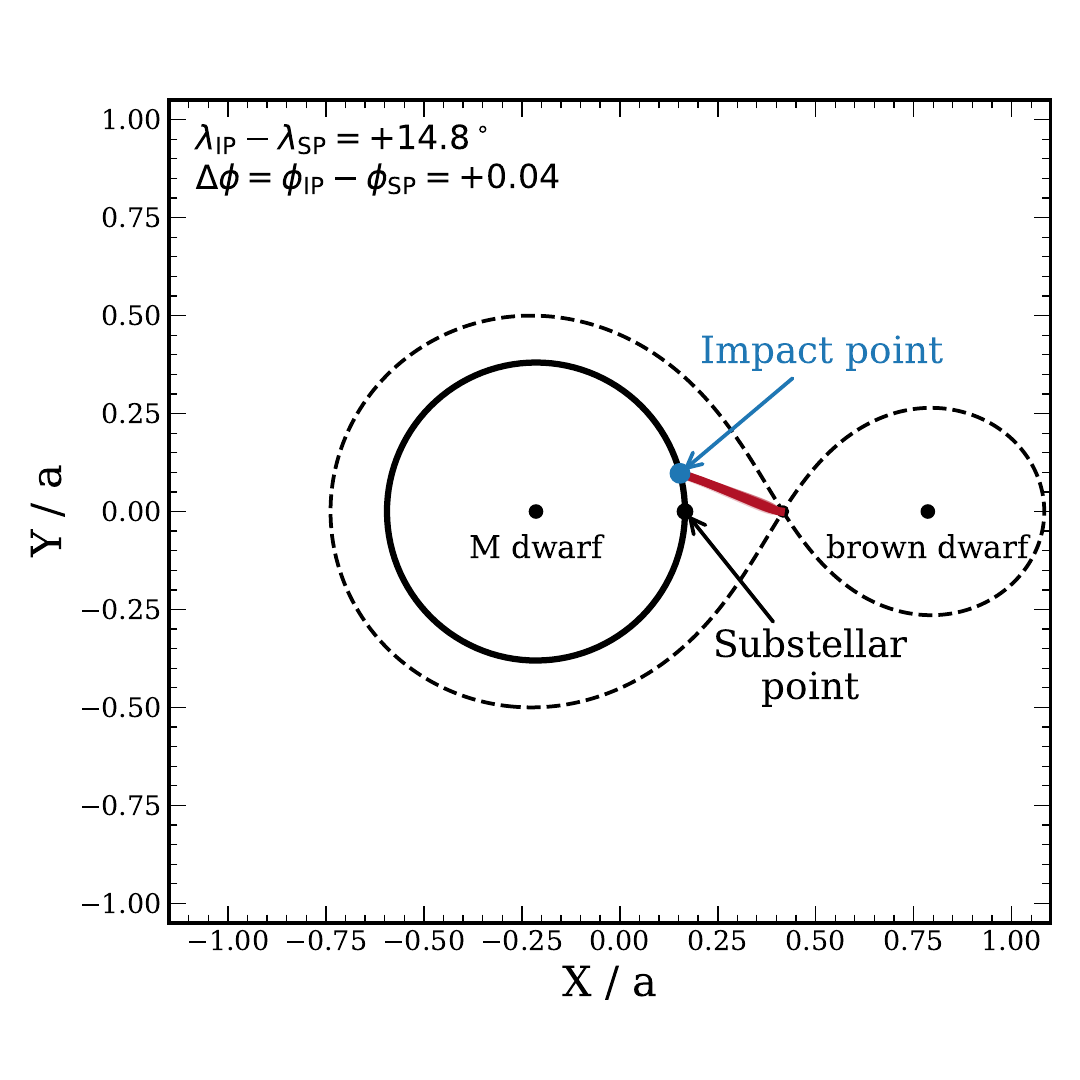}
    \caption{\textbf{Ballistic stream trajectories of test particles in the co-rotating frame of \Jzero.} The black points mark the centers of mass of the two components and the solid black circle shows the surface of the accretor. The dashed black curves indicate Roche-lobe equipotential surfaces, with the saddle point marking the inner Lagrange point (L1). The red curve shows trajectories of test particles launched from near L1, which directly impact the accretor surface rather than forming an accretion disk. The impact point (IP) is offset from the substellar point (SP) on the accretor by $\lambda_{\mathrm{IP}}-\lambda_{\mathrm{SP}} = 14.8^\circ$, corresponding to a phase offset $\Delta\phi \simeq 0.04$, consistent with the observed phase of maximum brightness in the HiPERCAM data.
    }
    \label{fig:schematic}
\end{figure}

In addition to this stability argument, the observations and the Roche-lobe ballistic stream geometry also point to the M dwarf being the accretor for \Jzero.  The ballistic stream model calculated in Figure~\ref{fig:schematic} indicates that the accretion stream impacts the accretor on its day side via direct-impact accretion. If this is the case, then the peak optical brightness from the hot spot should occur near superior conjunction of the accretor (i.e., at $\phi = 0.5$), when the day side of the accretor is most directly in view. Consistent with this prediction, the HiPERCAM light curves of \Jzero\ reach peak brightness at orbital phases $\phi\simeq0.44$--$0.48$, which is near superior conjunction of the M dwarf (as defined by the LRIS and ESI radial velocities of the M dwarf). Furthermore, the slight offset from superior conjunction closely matches the $\Delta\phi \simeq 0.04$ offset expected from Coriolis deflection of the ballistic stream before impacting the M dwarf (Figure~\ref{fig:schematic}, Methods), providing strong evidence that the M dwarf is the accretor in \Jzero.  Future phase-resolved radial velocities of the M dwarf in \Jone\ could provide an analogous test by tying the phase of peak brightness to the radial velocity of the M dwarf.

One of the most interesting features of the accretion in these systems is the large size of the hot spots. Although the ballistic accretion stream is expected to impact the accretor over a small localized region (Figure~\ref{fig:schematic}), fits to the LRIS blue-arm spectra near peak brightness imply hot-spot radii of order $R_{\rm hot}\sim10^{-2}\,R_\odot$ for both systems (Methods). The wavelength-dependent phase of maximum brightness in \Jzero\ (Figure~\ref{fig:J0440}) is also indicative of a large hot spot with a temperature gradient, where the cooler outer regions rotate into view before the hotter regions. Such large hot spots may arise if the thermal energy deposited by the accretion stream is efficiently redistributed (e.g., by strong winds on the accretor). Similar circulation-driven heat redistribution is seen in models of many irradiated atmospheres (e.g., for hot Jupiters \cite{Showman2002}), though quantitative modeling of energy transport in direct-impact accretion onto low-mass stars is largely unexplored to date. The inferred hot-spot sizes and wavelength-dependent phase shifts therefore imply that direct-impact accretion onto low-mass stars can produce complex and atypical hot spots that are far larger than those normally observed (e.g., in white dwarf accretors \cite{Cropper1990}), opening a new parameter space for hot-spot modeling (Methods).

In summary, \Jzero\ and \Jone\ are a new class of objects in which a Roche-lobe--filling brown dwarf transfers mass onto an M dwarf via direct-impact accretion. Together, these systems demonstrate that the fate of some substellar companions to main-sequence stars is not rapid engulfment and destruction, but instead long-lived stable mass transfer. Although theory has long allowed for such a configuration \cite{Rappaport1983,Hjellming1987}, it has not previously been observed. The discovery of \Jzero\ and \Jone\ therefore establishes a new evolutionary channel for substellar objects and raises the question of how such systems form. Because neither component in these binaries is evolved, \Jzero\ and \Jone\ could not have formed through common-envelope evolution, and instead likely formed at wider separations and were gradually driven inward until Roche-lobe overflow began. Indeed, there are known short-period M dwarf--brown-dwarf binaries which have separations only slightly wider than required for Roche-lobe overflow that are expected to reach contact within a few Gyr \cite{El-Badry2022,ElBadry2023}. However, the physical mechanism responsible for driving such low-mass binaries to such ultrashort orbital periods is not completely understood. Ultimately, addressing the question of how these systems formed likely requires moving beyond individual systems to a larger population. Fortunately, the \textit{Legacy Survey of Space and Time} (\textit{LSST}) at the Vera C. Rubin Observatory should be particularly effective at identifying more of these systems due to their strong chromatic variability and the six-band wavelength coverage of \textit{LSST}.

%%%%%%%%%%%%%%%% MAIN TEXT FIGURES %%%%%%%%%%%%%%%

%%%%%%%%%%%%%%%% MAIN TEXT TABLE %%%%%%%%%%%%%%%

\begin{table*}
\centering
\caption{Observational and physical properties of \Jone\ and \Jzero. The method used to infer each parameter is indicated in parentheses: ZTF = Zwicky Transient Facility; RV = radial velocities; LC = light-curve modeling; MCMC = Roche-geometry MCMC framework; SED/spectral fitting = \textit{SPHEREx} and LRIS fitting; Swift = \textit{Swift}/XRT.}
\footnotesize
\begin{tabular}{lcc}
\toprule
\textbf{Parameter} &
\textbf{\Jone} &
\textbf{\Jzero} \\
\midrule

Right Ascension (deg; ZTF)
& $221.0388$
& $70.0972$ \\

Declination (deg; ZTF)
& $+48.3456$
& $+23.4264$ \\

Orbital period (min; ZTF)
& $67.16010 \pm 8\times10^{-5}$
& $86.6467 \pm 1\times10^{-4}$ \\

Accretor mass $M_{\rm acc}$ ($M_\odot$; MCMC)
& $0.099^{+0.015}_{-0.016}$
& $0.106^{+0.017}_{-0.020}$ \\

Accretor mass $M_{\rm acc}$ ($M_{\rm Jup}$)
& $103^{+16}_{-16}$
& $111^{+17}_{-21}$ \\

Accretor radius $R_{\rm acc}$ ($R_\odot$)
& $0.119^{+0.015}_{-0.016}$
& $0.126^{+0.017}_{-0.020}$ \\

Accretor radius $R_{\rm acc}$ ($R_{\rm Jup}$)
& $1.18^{+0.15}_{-0.16}$
& $1.26^{+0.17}_{-0.20}$ \\

Accretor density $\rho_{\rm acc}$ (g\,cm$^{-3}$)
& $83^{+26}_{-16}$
& $74^{+28}_{-15}$ \\

Donor mass $M_{\rm don}$ ($M_\odot$; MCMC)
& $0.0397^{+0.0022}_{-0.0013}$
& $0.0288^{+0.0014}_{-0.0008}$ \\

Donor mass $M_{\rm don}$ ($M_{\rm Jup}$)
& $41.6^{+2.3}_{-1.3}$
& $30.2^{+1.5}_{-0.9}$ \\

Donor radius $R_{\rm don}$ ($R_\odot$; MCMC)
& $0.0856^{+0.0016}_{-0.0009}$
& $0.0907^{+0.0015}_{-0.0008}$ \\

Donor radius $R_{\rm don}$ ($R_{\rm Jup}$)
& $0.853^{+0.016}_{-0.009}$
& $0.903^{+0.015}_{-0.008}$ \\

Donor density $\rho_{\rm don}$ (g\,cm$^{-3}$)
& $89.1^{+2.3}_{-2.2}$
& $54.4^{+1.3}_{-1.3}$ \\

Inclination $i$ (deg; LC / RV+MCMC)
& $\gtrsim75$
& $24.4^{+3.6}_{-3.4}$ \\

Age (Gyr; MCMC)
& $5.2^{+3.2}_{-2.8}$
& $5.4^{+3.2}_{-3.0}$ \\

Semi-major axis $a$ (AU; derived)
& $(1.31^{+0.05}_{-0.05})\times10^{-3}$
& $(1.54^{+0.06}_{-0.08})\times10^{-3}$ \\

Roche-lobe radius $R_{L_{\rm Acc}}$ ($R_\odot$; MCMC)
& $0.129^{+0.008}_{-0.009}$
& $0.164^{+0.011}_{-0.014}$ \\

Roche-lobe radius $R_{L_{\rm Don}}$ ($R_\odot$; MCMC)
& $0.0856^{+0.0016}_{-0.0010}$
& $0.0907^{+0.0015}_{-0.0009}$ \\

Accretor filling factor $R_{\rm acc}/R_{L_{\rm Acc}}$ (MCMC)
& $0.918^{+0.057}_{-0.065}$
& $0.773^{+0.048}_{-0.063}$ \\

Donor filling factor $R_{\rm don}/R_{L_{\rm Don}}$ (MCMC)
& $1.000^{+0.008}_{-0.008}$
& $1.000^{+0.008}_{-0.008}$ \\

Density ratio $\rho_{\rm acc}/\rho_{\rm don}$
& $0.93^{+0.30}_{-0.19}$
& $1.36^{+0.52}_{-0.28}$ \\

M dwarf RV semi-amplitude $K$ (km\,s$^{-1}$; RV)
& ---
& $24.7 \pm 2.6$ \\

Distance (pc; SED/spectral fitting)
& $\sim200$--$300$
& $\sim100$--$200$ \\

X-ray luminosity $L_X$ (erg\,s$^{-1}$; Swift)
& $<7.6\times10^{29}$ (3$\sigma$) at 300 pc
& $<7.1\times10^{29}$ (3$\sigma$) at 200 pc \\

\bottomrule
\end{tabular}
\label{table1}
\end{table*}

\clearpage

%%%%%%%%%%%%%%%% METHODS %%%%%%%%%%%%%%%

\section*{Methods}
\label{Methods}
\textbf{Discovery in ZTF data}
We identified \Jzero\ and \Jone\ as part of a systematic search for periodic sources in data from the Zwicky Transient Facility (ZTF) \cite{Bellm2019,Graham2019,Masci2019}, which conducts a wide-field survey of the northern sky with $30$ second exposures in $g$, $r$, and $i$ bands. Following \cite{Burdge2020b}, we made use of forced photometry from ZTF difference images to create time series for approximately $1.3$ billion Pan-STARRS sources that satisfied simple color and magnitude cuts. We searched the combined $g$- and $r$-band light curves of all these sources using a GPU-accelerated implementation of the Lomb-Scargle periodogram with the trial frequency grid constructed between $f_{\mathrm{min}} = \frac{4}{1965}\,\mathrm{day}^{-1}$
and $f_{\mathrm{max}} = 360\,\mathrm{day}^{-1}$. We visually inspected the highest-significance short-period candidates that our search found. Most sources showed smooth or low-amplitude variability, but \Jone\ and \Jzero\ stood out due to their large-amplitude optical variability, which motivated targeted ground-based follow-up.

\begin{figure}
    \centering
    \includegraphics[width=\linewidth]{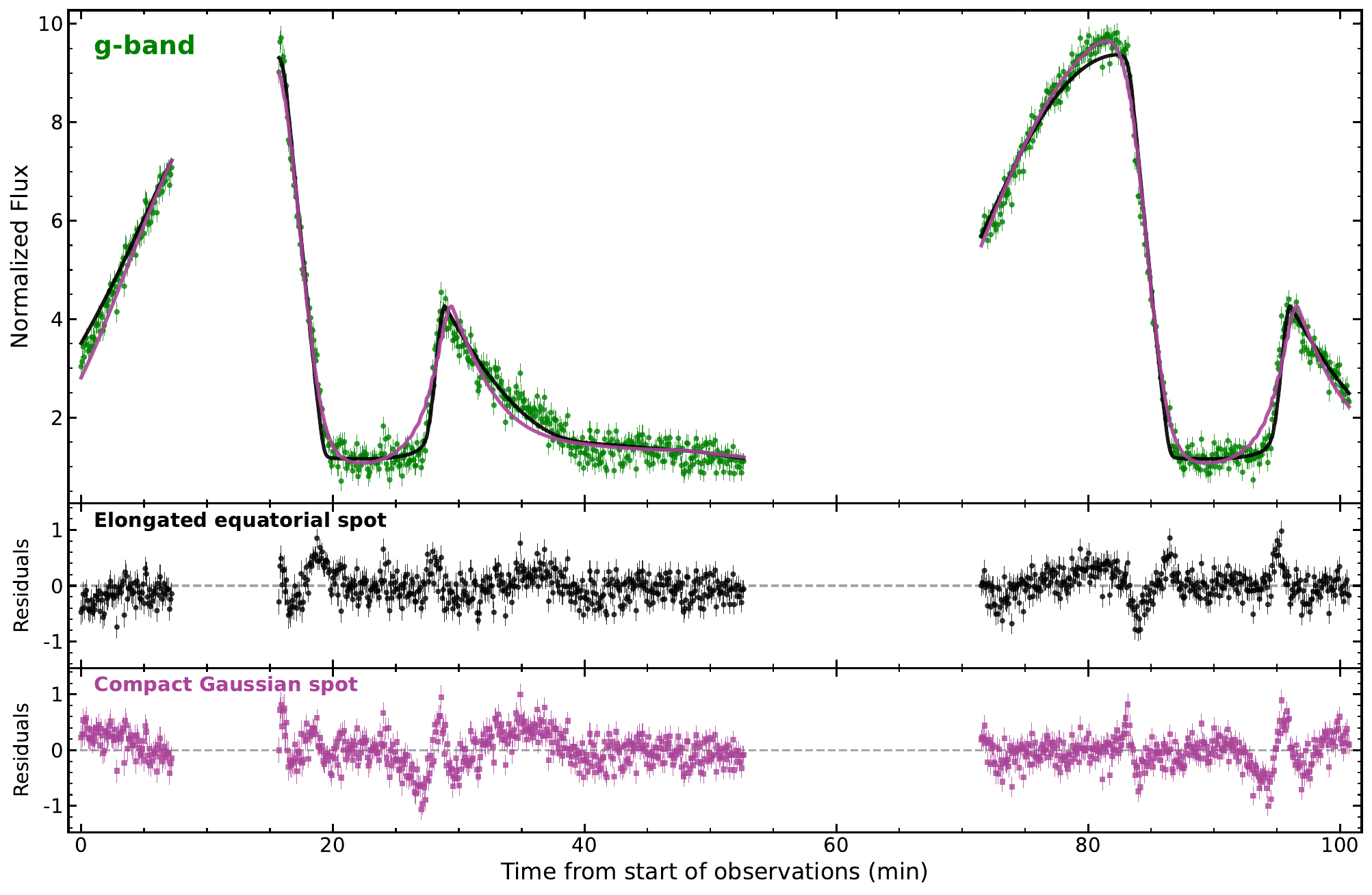}
    \caption{\textbf{Light-curve model fits for \Jone.}
    HiPERCAM $g$-band light curve of \Jone\ (green points with error bars) compared to two \texttt{lcurve} models (black and purple curves). The lower panel shows the residuals (data minus model) for each fit.
    }
    \label{fig:J1444lcurve}
\end{figure}

\textbf{HiPERCAM data}
We obtained high-speed optical photometry of \Jzero\ using HiPERCAM \cite{Dhillon2021} on UT 2023 October 18, with observations taken simultaneously in the $u$, $g$, $r$, $i$, and $z$ bands. \Jone\ was observed in the same configuration on UT 2024 August 3. The observations for \Jone\ were split into three blocks due to intermittent observing interruptions, but altogether they covered the majority of two orbital periods. We reduced each observing run independently using the standard HiPERCAM reduction pipeline \cite{Dhillon2021}. We performed aperture photometry on \Jzero\ and \Jone\ and on a nearby comparison star in each band, and constructed differential light curves by dividing the target counts by the comparison-star counts. All timestamps were converted from Modified Julian Date (MJD) to Barycentric Julian Date (BJD) in the TDB standard using the target coordinates and observatory location, and the resulting light curves were phase-folded on the orbital periods determined from the ZTF data. For \Jzero, the absolute orbital phase was anchored using the Keck/LRIS radial velocities. For \Jone, we simply  defined the eclipse midpoint from our equatorial spot \texttt{lcurve} fit (Extended Data Figure~\ref{fig:J1444lcurve}) to occur at $\phi=0.5$, so that $\phi=0.5$ corresponded to superior conjunction of the accretor in both systems. Following $3\sigma$ outlier rejection, each band was then normalized by its 5th-percentile flux to compare the variability amplitudes across wavelengths.

\textbf{Phase-resolved spectroscopy}
In addition to the photometric follow-up with HiPERCAM, we obtained phase-resolved optical spectroscopy of \Jzero\ and \Jone\ using the Low Resolution Imaging Spectrometer (LRIS) on the Keck I telescope. \Jzero\ was observed on UT 2022 February 2 using both arms of LRIS with the $5600\,\text{\AA}$ dichroic, the $600/4000$ grism on the blue arm, and the $400/8500$ grating on the red arm. \Jone\ was observed on UT 2024 June 28 using the blue arm with the $600/4000$ grism and a $1.0''$ slit. Unfortunately, no usable red-arm spectra were obtained for \Jone\ due to a focusing issue. All LRIS data were reduced using the LPipe pipeline \cite{Perley2019}, which performs bias subtraction, flat-fielding, cosmic-ray removal, spectral extraction, wavelength calibration, sky subtraction, and flux calibration using nightly standard stars. For \Jzero, we also obtained a second round of phase-resolved optical spectroscopy with the Echellette Spectrograph and Imager (ESI) on the Keck II telescope \cite{Sheinis2002} on UT 2022 September 1. The ESI data were reduced using the standard MAKEE pipeline (\url{sites.astro.caltech.edu/~tb/makee/}).

\textbf{Swift data}
Both \Jzero\ and \Jone\ were observed with the \textit{Neil Gehrels Swift Observatory} using the X-Ray Telescope (XRT). \Jzero\ was observed on UT 2021 December 8 and UT 2021 December 10 in two pointings and \Jone\ was observed on UT 2022 July 22 in three separate observations. Neither source was detected in X-rays, so we measured upper limits using the Leicester Swift XRT Point Source catalogue \cite{Evans2023}. For \Jzero, the two observations were combined into a stacked image with an effective exposure of $1.904\,\mathrm{ks}$, yielding a $3\sigma$ upper limit on the total-band ($0.3$--$10\,\mathrm{keV}$) count rate of $3.8\times10^{-3}\,\mathrm{s^{-1}}$.
For \Jone, the three observations were combined with an effective exposure of $4.012\,\mathrm{ks}$, yielding a $3\sigma$ upper limit of $1.8\times10^{-3}\,\mathrm{s^{-1}}$.
We converted these upper limits to $0.3$--$10\,\mathrm{keV}$ flux limits using WebPIMMS, assuming an absorbed power-law spectrum with photon index $\Gamma = 2$ and a Galactic hydrogen column density of $N_{\mathrm{H}} = 2 \times 10^{20}\,\mathrm{cm^{-2}}$. This yields flux upper limits of
$F_X < 1.5\times10^{-13}\,\mathrm{erg\,cm^{-2}\,s^{-1}}$ for \Jzero\ and
$F_X < 7.0\times10^{-14}\,\mathrm{erg\,cm^{-2}\,s^{-1}}$ for \Jone.
At a distance of $d=200\,\mathrm{pc}$ for \Jzero\ and $d=300\,\mathrm{pc}$  for \Jone, these correspond to
$L_X < 7.1\times10^{29}\,\mathrm{erg\,s^{-1}}$ and
$L_X < 7.6 \times10^{29}\,\mathrm{erg\,s^{-1}}$, respectively.

\textbf{Radial velocities in \Jzero}
We measured radial velocities for \Jzero\ from sixteen LRIS red-arm spectra and fourteen ESI spectra using the Na I doublet at $8183\,\text{\AA}$ and $8195\,\text{\AA}$. We fit the Na I doublet directly rather than using cross-correlation because these sodium absorption features are unambiguously photospheric features of the M dwarf, whereas other features in the spectra (e.g., the Balmer lines) are contaminated by the hot spot and could potentially bias the radial velocity measurements. For each spectrum, we normalized the local continuum following a similar iterative spline approach to \cite{Householder2025} and fit the Na I doublet with a double-Voigt profile in which both lines were constrained to have the same Doppler shift, while the line depths and widths were fit independently. Each fit was performed using weighted least-squares minimization, with per-pixel uncertainties propagated through the covariance matrix to estimate the uncertainty for each spectrum. While the individual Na I fits from ESI spectra had reduced chi-squared values near unity, a best-fitting sinusoidal model to the ESI radial-velocity time series yielded a reduced chi-squared much larger than unity, indicating that the uncertainties underestimated the true RV scatter. We therefore inflated the ESI uncertainties by adding a jitter term in quadrature,  $\sigma_{\rm total}=\sqrt{\sigma_{\rm fit}^2+\sigma_{\rm jitter}^2}$, chosen to give a reduced chi-squared of unity for the best-fit sinusoidal model to the ESI data.

To jointly model both the LRIS and ESI radial velocities, the observation times were converted from MJD to BJD (TDB), and the orbital phases were computed using the photometric ephemeris from the ZTF search. We then shifted the LRIS radial velocities to match the ESI systemic velocity in order to account for any instrument-dependent RV zero-point offsets. By eye, the joint radial velocities show a clear sinusoidal modulation (Figure~\ref{fig:J0440}d) at the $86.65$ minute orbital period, indicating that the measured radial velocities indeed track the orbital motion of the M dwarf. The combined radial velocities were fit with a sinusoidal Markov Chain Monte Carlo (MCMC) model assuming a circular orbit at the period recovered in ZTF, which yielded a semi-amplitude of $K = 24.7 \pm 2.6\,\mathrm{km\,s^{-1}}$. We also refined the orbital phasing by shifting the phase of the best-fit sinusoidal model and the observed radial velocities so that the maximum radial velocity of the model occurred at $\phi = 0.25$, which had an uncertainty of $\sigma(\phi) = 0.02$. With this convention, $\phi = 0.5$ corresponds to superior conjunction of the M dwarf, which allowed for a direct comparison between the radial velocities and the relative orbital phase of peak optical brightness using the ephemeris from ZTF.

\textbf{Mass and radius constraints}
To quantify the component masses, radii, and ages  self-consistently, we modeled each system using a MCMC framework that enforced Roche geometry for an M dwarf accretor and brown dwarf donor with physically motivated mass--radius relations. We allowed the accretor mass, $M_{\rm acc}$, donor mass, $M_{\rm don}$, and system age (sampled uniformly between $1$ Gyr and $10$ Gyr) to vary.  For \Jzero, we also fit the orbital inclination $i$ (sampled uniformly in $\cos i$), which was constrained by the measured radial-velocity semi-amplitude of the M dwarf. The donor radius was obtained from hydrogen-rich brown-dwarf evolutionary tracks computed with MESA \cite{Paxton2011} from \cite{Nelson2018}, interpolated from the mass and age with the multidimensional linear interpolation scheme in \texttt{scipy} \cite{scipy}. The donor was required to fill its Roche lobe at each MCMC step using the Eggleton analytic approximation \cite{Eggleton1983}, evaluated for the given mass ratio and orbital separation derived from Kepler's third law. This Roche-lobe filling requirement was enforced via an exponential penalty that sharply suppressed solutions in which the donor radius differed from the Roche-lobe radius by more than $1\%$.  For the accretors in \Jzero\ and \Jone, we adopted a simple power-law main-sequence M dwarf mass--radius relation,
\begin{equation}
\frac{R_{\rm acc}}{R_{\odot}} = 0.85 \left(\frac{M_{\rm acc}}{M_{\odot}}\right)^{0.85},
\label{MR}
\end{equation}
based on fits to theoretical lower main-sequence models \cite{NelsonPrivateComm2024}. Uniform priors were placed on the accretor mass, where we restricted $0.075 \le M_{\rm acc} \le 0.13\,M_{\odot}$ for \Jzero, given the empirical match with the M8 template in \texttt{PyHammer}  (Figure~\ref{fig:J0440}), and $0.075 \le M_{\rm acc} \le 0.2\,M_{\odot}$ for \Jone, reflecting the lack of strong constraints from the \textit{SPHEREx} data.  In both systems, we also required that the accretor underfilled its Roche lobe with the same Eggleton prescription.

With this setup, we sampled the posterior space using the affine-invariant ensemble sampler \texttt{emcee} \cite{Foreman-Mackey2013} with $32$ walkers and $5000$ steps, discarding the first $1000$ steps as burn-in. Convergence was verified by confirming that the Gelman-Rubin statistic ($\hat{R}$) satisfied $\hat{R} < 1.01$ for all fitted parameters.  The resulting posterior constraints on the component masses, radii, and ages are summarized in Table~\ref{table1} with all quoted uncertainties representing 68\% ($1\sigma$) confidence intervals unless otherwise stated.
However, it is very important to note that the reported uncertainties reflect the statistical posteriors within our adopted physical framework, but do not capture systematic uncertainties associated with the choice of evolutionary models or mass--radius prescriptions. We therefore caution that the quoted uncertainties should not be interpreted as absolute errors on the physical parameters, but rather as conditional uncertainties given the assumed underlying mass-radius prescriptions and evolution models. While it is possible to artificially inflate the reported error bars to account for model uncertainty, doing so requires subjective choices about the differences between various mass-radius prescriptions or more detailed analysis that is beyond the scope of this paper. We instead report the posterior uncertainties exactly as obtained from the MCMC sampling, and explicitly emphasize that systematic uncertainties associated with the mass--radius relation likely dominate the true error budget.

\textbf{Light-curve modeling}
We modeled the HiPERCAM light curves of both \Jzero\ and \Jone\ using \texttt{lcurve} \cite{Copperwheat2010}, which has the capability to simultaneously model both Roche-lobe--filling binaries and accretion hot spots. For \Jzero, we fixed the orbital inclination, component radii, and mass ratio to the median values from the Roche-geometry and radial-velocity constraints (Table~\ref{table1}). For \Jone, we adopted the median values from the Roche-geometry constraints and allowed the orbital inclination to vary. In both systems, we also fit parameters to characterize the accretion hot spot. The \texttt{lcurve} code implements two simplified spot prescriptions: a compact circular Gaussian spot, parameterized by an effective temperature and full width at half maximum at an arbitrary latitude and longitude, and an equatorial spot fixed at zero latitude, with adjustable longitudinal extent, thickness, and edge fall-off.

\begin{figure}[h]
    \centering
    \includegraphics[width=\linewidth]{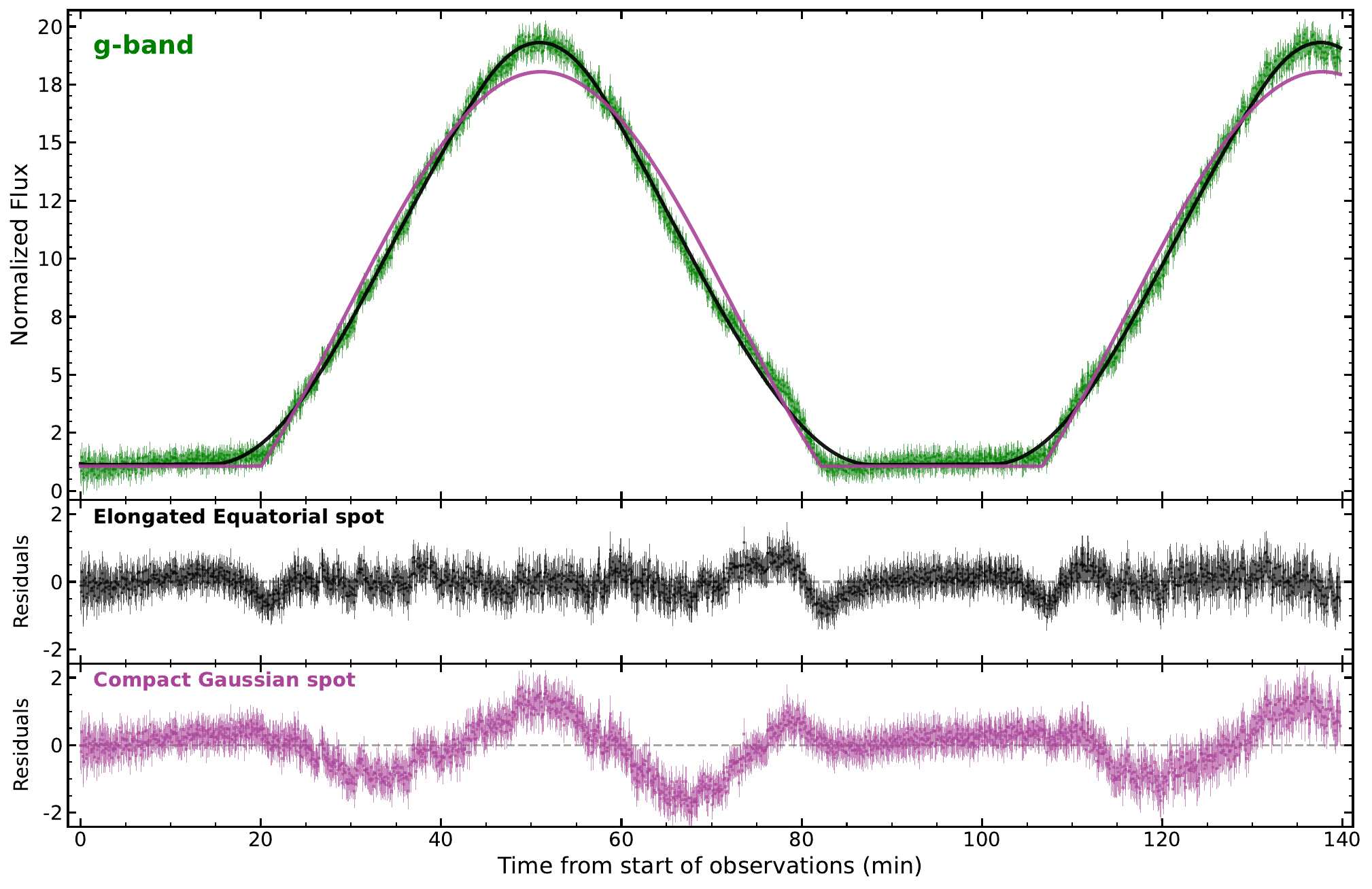}
    \caption{\textbf{Light-curve model fits for \Jzero.}
    HiPERCAM $g$-band photometry (green) compared to two different hot-spot models computed with \texttt{lcurve} (black and purple). The residuals (data minus model) are shown below the fits. Both models reproduce the overall variability but do not fully capture the detailed light-curve structure, indicating limitations of simplified spot prescriptions in reproducing the observed hot-spot structure.}
    \label{fig:J0440lcurve}
\end{figure}

Using these two spot prescriptions, we optimized the light-curve models using the Levenberg--Marquardt algorithm.  We also attempted to explore the full parameter space using a Markov Chain Monte Carlo (MCMC) framework, but the chains failed to converge for both systems due to the simplified treatment of spot physics in \texttt{lcurve}. In particular, the compact Gaussian-spot model reproduced the sharp ingress and egress observed in the data but underpredicted the flux near peak brightness, while the equatorial-spot model matched the overall flux distribution near peak brightness but produced ingress and egress that are significantly smoother than the observations (Extended Data Figures~\ref{fig:J0440lcurve} and \ref{fig:J1444lcurve}). These discrepancies suggest that the simplified spot geometries implemented in \texttt{lcurve} are insufficient to capture the true complex structure of the hot spots in these systems. Indeed, the phase-dependent peak brightness in Figure~\ref{fig:J0440} indicates that these hot spots have complex structure, likely consisting of a compact impact point that extends into a larger, elongated region with a temperature and brightness gradient. Because such complex spot configurations are not implemented in \texttt{lcurve}, we emphasize that our light-curve models are intended as qualitative demonstrations that the observed variability can be produced by hot spots arising from direct-impact accretion, rather than as physically complete descriptions of the hot spots.

In addition to this \texttt{lcurve} modeling, we also quantified the phase of maximum brightness in \Jzero\ by fitting the HiPERCAM light curves with a fifth-order Fourier series, $f(\phi)=a_0+a_1\cos(2\pi\phi)+b_1\sin(2\pi\phi)+a_2\cos(4\pi\phi)+b_2\sin(4\pi\phi)$, using weighted least squares.  The phase of peak brightness was then obtained by numerically maximizing this fit.  Using the tightly constrained orbital period from ZTF, we found that the maximum brightness is offset from superior conjunction of the M dwarf in the radial velocities by $\phi = 0.02$--$0.06$, depending on the wavelength. We do not measure an analogous phase offset in \Jone, since no radial velocities exist to define an absolute phase reference.

\textbf{Ballistic stream model}
To provide a physically motivated interpretation of the phase offset in \Jzero, we constructed a simple ballistic model of the accretion stream in the co-rotating frame of the two objects. We treated the donor and accretor as point masses on a circular orbit and tracked the motion of test particles that represented parcels of gas leaving the donor. All lengths were expressed in units of the orbital separation $a$, and we adopted dimensionless units with $M_{\mathrm{tot}}=1$ and $\Omega=1$, so that time is measured in units of $\Omega^{-1}$ and velocities in units of $a\Omega$. Under this normalization, in the rotating frame, the effective Roche potential is given by
\begin{equation}
\Phi(x,y) = -\frac{1-\mu}{r_1} - \frac{\mu}{r_2}
            - \frac{1}{2}\left(x^2 + y^2\right),
\end{equation}
where $\mu = M_{\mathrm{don}}/(M_{\mathrm{acc}}+M_{\mathrm{don}})$, $r_1=\sqrt{(x+\mu)^2+y^2}$, and $r_2=\sqrt{(x-(1-\mu))^2+y^2}$. In this coordinate system, the accretor and donor are located at $x_{\mathrm{acc}}=-\mu$ and $x_{\mathrm{don}}=1-\mu$, respectively. The inner Lagrange point (L1) is defined by $\partial\Phi/\partial x=0$ at $y=0$, and the Roche-lobe geometry shown in Figure~\ref{fig:schematic} corresponds to the equipotential surface passing through L1.

To connect this Roche geometry to the photometric behavior of \Jzero, we modeled the trajectory of gas leaving the donor and impacting the accretor using the median parameters from our posterior distributions. Gas leaving the donor was represented by massless test particles launched from near L1 and integrated forward in time using a second-order leapfrog scheme with timestep $\Delta t = 2\times10^{-3}$. The initial conditions of the test particles were chosen to reflect the physics of Roche-lobe overflow near L1. As shown by \cite{Lubow1975}, gas escapes through a narrow nozzle in the effective potential, with the stream thickness set by the local sound speed rather than the orbital velocity. We therefore parameterized the launch velocities using the dimensionless thermal speed
\begin{equation}
\epsilon = \frac{c_s}{\Omega a},
\end{equation}
where $c_s$ is the gas sound speed. We adopted a gas temperature of $T=3000\,\mathrm{K}$, and initialized the ballistic trajectory just downstream of L1 and verified that varying the gas temperature by factors of a few, or modestly shifting the launch location, did not meaningfully affect the results of the integration.

In the rotating frame, the Coriolis force rapidly focuses the trajectories of the test particles, causing the launch paths to converge onto nearly the same path before reaching the accretor (Figure~\ref{fig:schematic}). Importantly, the stream does not circularize outside the accretor, so the accretion occurs via direct-impact rather than through a disk. To quantify where the stream impacts the accretor, we defined the substellar point (SP) as the point on the accretor directly facing the donor, and the impact point (IP) as the location where the ballistic stream first intersects the accretor surface. The ballistic calculation gives a longitudinal offset on the accretor surface of $\lambda_{\mathrm{IP}}-\lambda_{\mathrm{SP}}\simeq 14.8^\circ$, implying a phase offset $\Delta\phi=(\lambda_{\mathrm{IP}}-\lambda_{\mathrm{SP}})/360^\circ \simeq 0.04$. Importantly, because $\lambda_{\mathrm{IP}}-\lambda_{\mathrm{SP}}>0$, the hot spot lies ahead of the substellar point in the direction of orbital motion, so it rotates into view before the substellar point by $\phi\approx0.04$, consistent with the observations in Figure~\ref{fig:J0440}.

\textbf{Stability of mass transfer}
In direct-impact accretion, gas leaving the Roche-lobe--filling donor carries orbital angular momentum and deposits it into the spin of the accretor when it strikes its surface. In the absence of any additional torques, this would spin up the accretor and remove angular momentum from the orbit, leading to runaway orbital inspiral and unstable mass transfer. However, because the spin--orbit synchronization timescale, $t_{\rm sync}$, scales strongly with orbital separation as $t_{\rm sync}\propto (a/R)^6$ \cite{Hurley2002}, the tidal torques are expected to be extremely strong at these close orbital separations (Table \ref{table1}). Thus, even modest deviations from co-rotation are likely rapidly damped by tides. As a result, any angular momentum deposited into the spin of the accretor is likely efficiently transferred back into the orbit through tides, leading to long-lived mass transfer rather than runaway spin-up of the accretor.

The stability of Roche--lobe overflow has also been investigated more quantitatively (e.g., \cite{Rappaport1983,Hjellming1987}). In particular, Equation 33 of \cite{Rappaport1983} introduced a stability criterion for Roche lobe filling donors, where both the numerator and denominator of this equation must be positive for stable mass transfer. Importantly, this criterion provides a quantitative way to evaluate which binary configurations are compatible with stable mass transfer (e.g., in terms of mass ratio). For conservative stable mass transfer (i.e., all the mass that leaves the donor is accreted by the companion), the denominator (D) in their Equation 33 simplifies to
\begin{equation}
\label{stability}
D=\frac{5}{6}+\frac{\xi_{\rm ad}}{2}-\frac{M_{\rm don}}{M_{\rm acc}},
\end{equation}
where $\xi_{\rm ad}=d\ln R_{\rm don}/d\ln M_{\rm don}$ is the adiabatic response of the radius of the donor to mass loss. For hydrogen-rich brown dwarfs and fully convective low-mass M dwarfs, $\xi_{\rm ad}\simeq -1/3$ \cite{Hjellming1987,Chabrier2000}, which implies that removing mass causes the donors to expand on dynamical timescales. Substituting $\xi_{\rm ad}\simeq -1/3$ into Eq.~\ref{stability} yields $D\simeq\tfrac{2}{3}-M_{\rm don}/M_{\rm acc}$, implying that stable mass transfer requires $M_{\rm don}/M_{\rm acc}\lesssim 2/3$. For fully convective donors, detailed calculations also give a similar mass ratio requirement, $q<0.634$ \cite{Hjellming1987}, which indicates that the more massive M dwarf is the accretor in these systems rather than the brown dwarf. Importantly, the posterior mass ratios, $q = 0.27^{+0.06}_{-0.04}$ for \Jzero\ and $q = 0.41^{+0.07}_{-0.06}$ for \Jone, are both comfortably below this threshold for stable mass transfer.

These results indicate that the mass transfer in \Jzero\ and \Jone\ is likely dynamically stable, but they do not constrain how long this stability lasts. Quantifying the precise lifetime of this mass transfer would require detailed modeling of the coupled mass-transfer, tidal, and angular-momentum evolution in these systems, which is beyond the scope of this work. Nonetheless, an empirical argument can be made. The most plausible route to Roche-lobe contact in these systems is long-term angular-momentum loss operating over Gyr timescales \cite{ElBadry2023}, and very few non-interacting M dwarf--brown dwarf binaries are known at such short orbital periods \cite{ElBadry2023}. It would therefore be unlikely to observe two systems that had only just initiated contact if the ensuing mass-transfer phase were short-lived or rapidly unstable. We therefore conclude that the mass transfer in \Jzero\ and \Jone\ is likely dynamically stable for at least $\gtrsim 10^{7}$--$10^{8}$ yr, and plausibly hundreds of Myr to several Gyr, although more detailed population studies and evolutionary modeling will be required to quantify this further.

\textbf{Spectral modeling and distances}
\Jzero\ does not have a Gaia parallax measurement. We therefore estimated its distance photometrically by comparing our flux-calibrated LRIS spectrum near minimum optical brightness (when the contribution from the hot spot is least) to BT--Settl atmosphere models \cite{Allard2012}. We used the BT--Settl grid for $T_{\rm eff}=1000$--$4000$ K based on the solar abundance pattern of \cite{Asplund2009}, limiting the  metallicities to ${\rm [M/H]} > -1.0$ and surface gravities to $\log g = 5.0$--$5.5$. We degraded each of the models to match the resolution of LRIS ($R\simeq 2000$) using Gaussian convolution. For each model characterized by effective temperature ($T_{\mathrm{eff}}$),  $\log g$ and [M/H], we estimated the stellar radius using the empirical $T_{\mathrm{eff}}$--radius relations of \cite{Rabus2019}. We also required the stellar radii implied by this relationship to lie within the $3\sigma$ MCMC posterior distribution from our Roche-lobe geometry constraints.

\begin{figure}
    \centering
    \includegraphics[width=\linewidth]{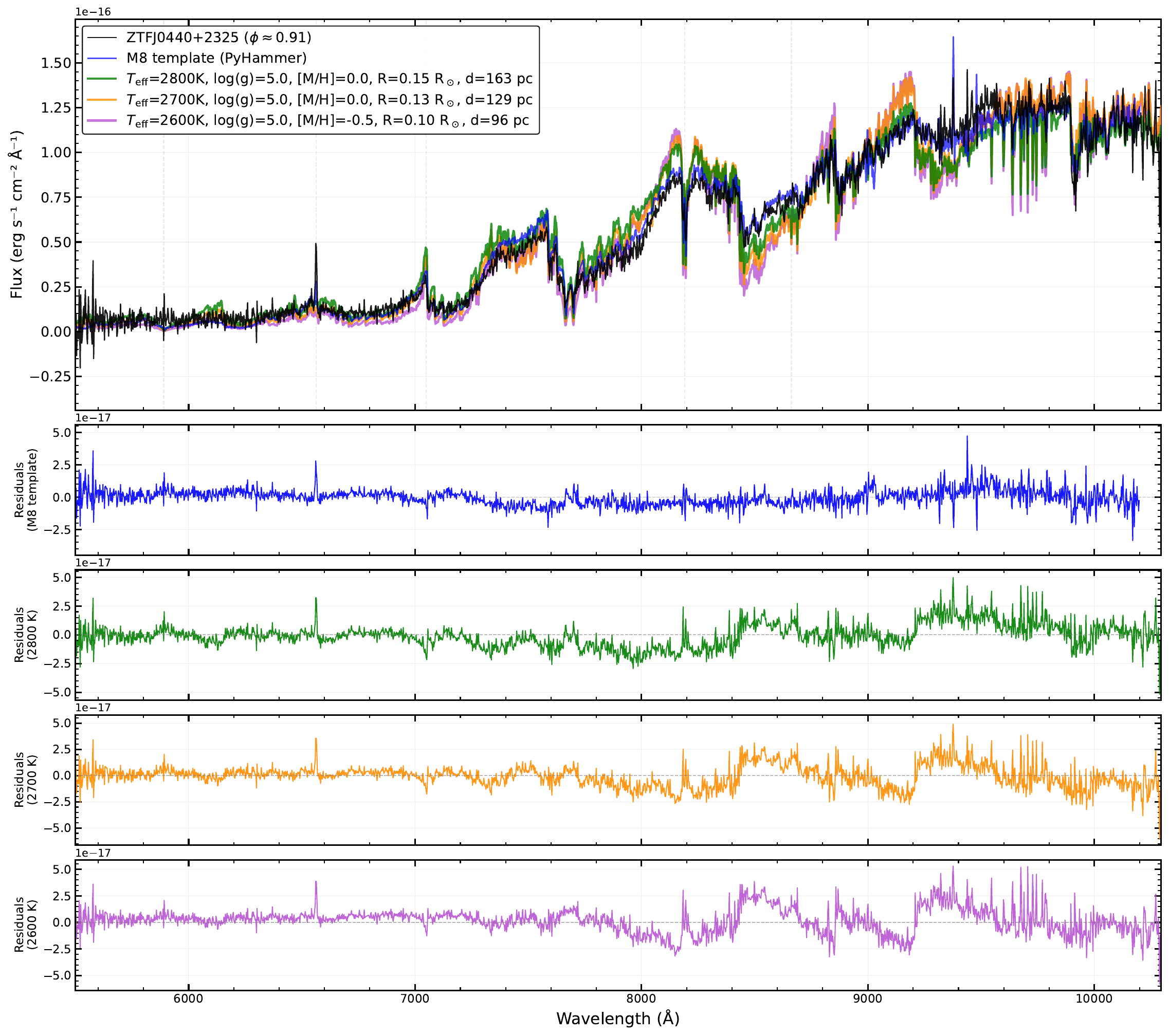}
    \caption{\textbf{Atmosphere model fits for \Jzero.}
    In the top panel, a flux-calibrated Keck/LRIS optical spectrum obtained near minimum brightness (black) is compared to an empirical M8 dwarf template and BT--Settl atmosphere models (colored lines). The BT--Settl fits imply a photometric distance of approximately $100$--$200$ pc. Residuals are shown below each fit.}
    \label{fig:J0440BT--Settl}
\end{figure}

The observed spectrum was then fit to the model spectrum via the scaling relation, $
F_{\mathrm{obs}} = F_{\mathrm{model}}\left(\frac{R_{\star}}{d}\right)^2,$
from which we inferred the distance. To evaluate the performance of the fits, we binned the LRIS spectra in wavelength so that each bin corresponded to approximately one instrumental resolution element, before computing the chi-squared ($\chi^2$) statistic. The resulting $\chi^2$ values were used only as a heuristic metric to rank models within the grid, rather than as a formal goodness-of-fit statistic, because the residuals for the fits still had significant correlated structure (likely due to the limitations of the BT--Settl models). The three best-fitting models corresponded to late-M dwarf atmospheres with $T_{\rm eff}= 2600$--$2800\,\mathrm{K}$, which imply distances of order $100$--$200\,\mathrm{pc}$ (Extended Data Figure~\ref{fig:J0440BT--Settl}). For comparison, we also fit the observed spectrum to empirical late-M dwarf templates from \texttt{PyHammer}. Because the \texttt{PyHammer} templates are defined by spectral type rather than effective temperature, we did not use them to infer a distance and instead fit only for an arbitrary flux scaling for visual comparison. The scaled M8 \texttt{PyHammer} template provided an almost exact match to the observed molecular absorption features (Figure~\ref{fig:J0440} and Extended Data Figure~\ref{fig:J0440BT--Settl}), so we therefore classified the accretor in \Jzero\ as an M8 dwarf.

\begin{figure}
    \centering
    \includegraphics[width=\linewidth]{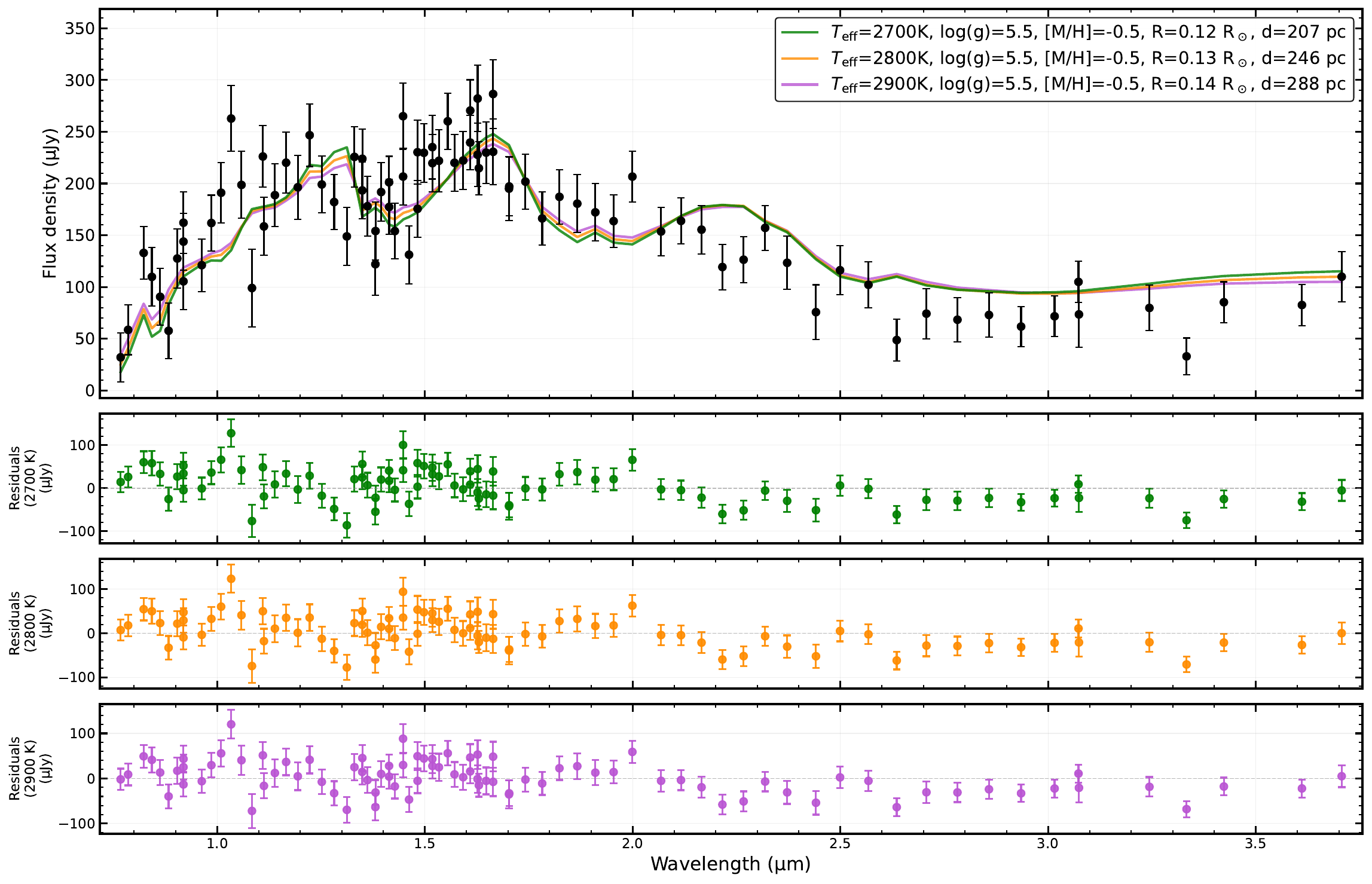}
    \caption{\textbf{SPHEREx spectral fitting for \Jone.}
    \textit{SPHEREx} near-infrared spectro-photometry of \Jone\ (black points with error bars) compared to BT--Settl atmosphere models with effective temperatures between $2700$ and $2900\,\mathrm{K}$ (colored lines). Residuals for each model are shown in the lower panels.}
    \label{fig:spherex}
\end{figure}

For \Jone, Gaia astrometry yields a nominal distance of $663\pm 245\,\mathrm{pc}$. However, the parallax has low significance and the astrometric solution reports significant excess noise ($2.5$ mas at a significance of 2.9), so we treat the Gaia distance as uncertain and potentially biased due to the strong variability and do not rely on it for our physical interpretation. Instead, we analyzed the publicly available flux-calibrated \textit{SPHEREx} spectro-photometry from $0.75$--$3.75\,\mu\mathrm{m}$.  In particular, we quantified the spectral type and photometric distance of \Jone\ using the same approach outlined above for \Jzero. We fit only \textit{SPHEREx} bands 1--4, as the redder bands 5--6 were dominated by large uncertainties and occasional negative flux values and showed no clear spectral features. At the resolution of \textit{SPHEREx} ($R\sim40$--$130$), the data are best fit by a single late-type stellar photosphere, with M dwarf models of $T_{\mathrm{eff}}= 2700$--$2900\,\mathrm{K}$ at distances on the order of $200$-$300\,\mathrm{pc}$ providing the best fits to the data in terms of $\chi^2$ (Extended Data Figure~\ref{fig:spherex}). Importantly, no second component is required to reproduce the \textit{SPHEREx} data, which is consistent with a cool or cloudy brown-dwarf donor that does not produce strong infrared emission. Nevertheless, the \textit{SPHEREx} measurements are not phase--resolved and combine flux from different orbital phases at different wavelengths, so the inferred temperature of the accretor and distance to the system should be interpreted with some caution.  Definitive characterization of the component masses and temperatures will ultimately require phase-resolved spectroscopy in the near-infrared (as was done for \Jzero).

\begin{figure*}
    \label{fig:accretion_temperature}
    \centering
    \includegraphics[width=0.8\linewidth]{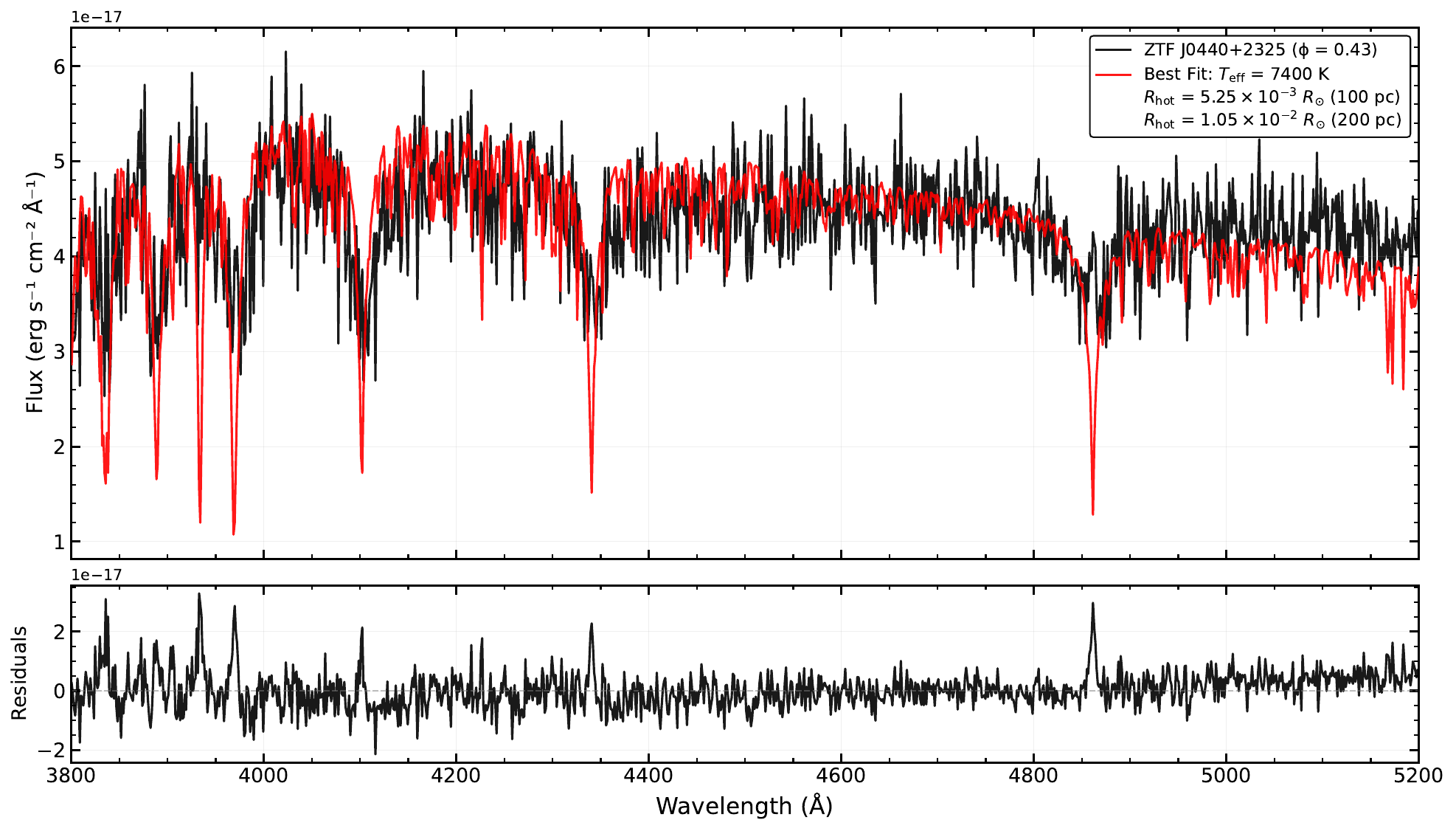}
    \includegraphics[width=0.8\linewidth]{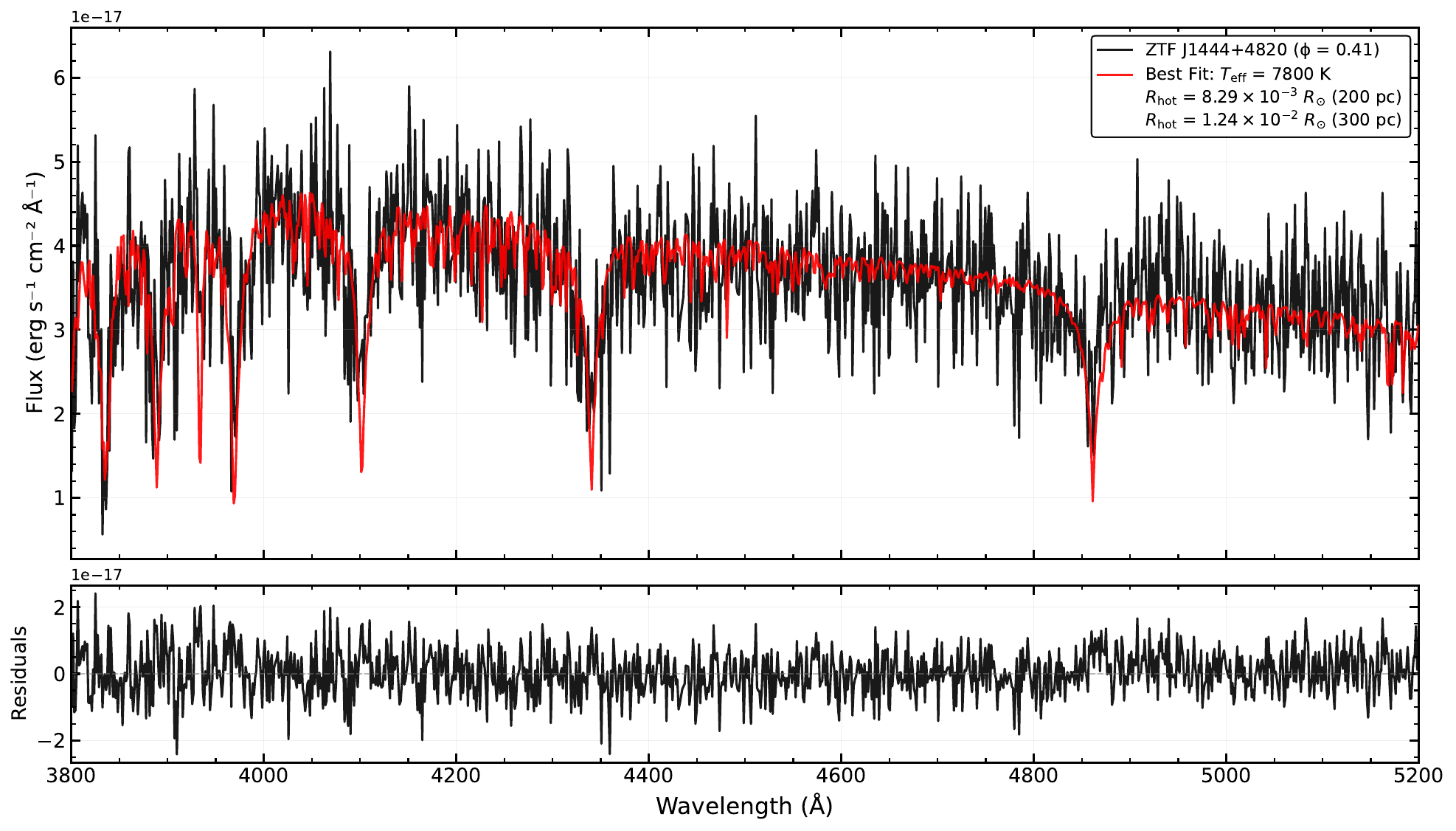}
    \caption{\textbf{Hot-spot temperature constraints.}
    Blue-arm spectra obtained near peak brightness compared to BT--Settl atmosphere models. The best-fitting models imply hot-spot temperatures of $\sim7400$ K for \Jzero\ and $\sim7800$ K for \Jone. Residuals are shown below each fit.}
\end{figure*}

\textbf{Estimating the hot-spot size}
To estimate the temperature of the hot spot, we modeled only LRIS blue-arm spectra obtained near peak orbital brightness, when the hot spot dominates the flux. We fit the spectra over $3800$--$5200\,\text{\AA}$ using solar-metallicity BT--Settl atmosphere models spanning $T_{\rm eff}=4000$--$10000$ K, which were degraded to the LRIS spectral resolution and scaled by a single multiplicative factor (exactly as was done for the M dwarf spectral fitting).  For \Jzero, the best-fitting model had an effective temperature of $T_{\rm eff}\sim7400$~K, while for \Jone\ we obtained $T_{\rm eff}\sim7800$~K (Extended Data Figure~\ref{fig:accretion_temperature}). Fits to spectra at nearby orbital phases that also have Balmer absorption features gave similar results, with modest differences at the level of a few hundred Kelvin, which we adopt as the representative uncertainty on the temperature. The scale factor obtained from these fits also allowed us to infer the angular size and radii of the hot spot ($R_{\rm hot}$). Because the distances to both systems are uncertain, we report hot-spot sizes for two different distances motivated by the photometric constraints discussed above. For \Jzero, the hot-spot radius is $R_{\rm hot}=5.25\times10^{-3}\,R_\odot$ at $d=100$~pc and $R_{\rm hot}=1.05\times10^{-2}\,R_\odot$ at $d=200$~pc. For \Jone, we find $R_{\rm hot}=8.20\times10^{-3}\,R_\odot$ at $d=200$ pc and $R_{\rm hot}=1.24\times10^{-2}\,R_\odot$ at $d=300$ pc.

\textbf{Compact-object alternatives}
Because both \Jzero\ and \Jone\ have ultrashort orbital periods and extreme chromatic variability, it is important to consider whether either system could instead host a compact object, such as in a polar or a black widow pulsar, rather than a brown dwarf donor.  Although the light-curve morphology of both systems is superficially similar to polars, this interpretation is strongly disfavored by the spectroscopy: neither system shows the cyclotron humps that normally dominate the optical spectra of polars. Furthermore, in both systems, the hot spot has a radius on the order of $R_{\rm hot}\sim10^{-2}\,R_\odot$, which is comparable to a typical white dwarf radius and several orders of magnitude larger than the localized accretion spots characteristic of polars \cite{Cropper1990}.

For \Jzero, the measured radial velocity semi-amplitude of the M dwarf ($K = 24.7 \pm 2.6\,\mathrm{km\,s^{-1}}$) also strongly disfavors both the polar and black widow pulsar interpretations.  For a circular binary, the semi-amplitude of the visible star is
\begin{equation}
K = \left(\frac{2\pi G}{P}\right)^{1/3}
\frac{M_{\rm acc}}{(M_{\rm acc}+M_{\rm don})^{2/3}}\sin i.
\end{equation}

\noindent If the accretor were a $0.6\,M_\odot$ white dwarf with a $0.11\,M_\odot$ M dwarf companion, an edge-on binary at $P=86.65$ minutes would give $K_{\rm edge}=409.7\,\mathrm{km\,s^{-1}}$ for the M dwarf, implying that the observed radial velocity semi-amplitude could only be matched for $i\le3.4^\circ$, which has an a priori probability of $1.8\times10^{-3}$ for a randomly oriented system. If the accretor were instead a neutron star, the inclination requirement is even more extreme. For a $1.4\,M_{\odot}$ neutron star and a $0.11\,M_{\odot}$ companion, the edge-on M dwarf radial velocity at this period would be $K_{\mathrm{edge}}=578.0\,\mathrm{km\,s^{-1}}$. This implies that a neutron-star-M dwarf binary would require $i\le2.5^\circ$ to match the observed radial velocities of the M dwarf, which has an a priori probability of $9 \times10^{-4}$.  Furthermore, \Jzero\ also closely resembles the published object ZTF J1406+1222, which was previously interpreted as a low-inclination black widow pulsar \cite{Burdge2022}. Both \Jzero\ and ZTF J1406+1222 exhibit large-amplitude optical variability, transient Balmer absorption that appears near peak brightness, and unexpectedly small radial velocity semi-amplitudes that are difficult to reconcile with a neutron-star companion. For ZTF J1406+1222, the inclination required to match the observed radial velocities ($K = 112 \pm 15\,\mathrm{km\,s^{-1}}$) corresponds to a probability of $1.5\times10^{-2}$ for a $1.4\,M_{\odot}$ neutron star and a $0.11\,M_{\odot}$ companion. The probability that both \Jzero\ and ZTF J1406+1222 are independently oriented at such low inclinations is therefore $1.4\times10^{-5}$, making the black widow explanation highly unlikely. Although \Jone\ lacks phase-resolved radial velocities, its sharp optical eclipses rule out a black widow pulsar interpretation, since black widow pulsar optical variability is driven purely by irradiation of the companion.

Furthermore, both of these compact-object scenarios also predict strong high-energy emission. Polars typically have X-ray luminosities $L_X\sim10^{31}$--$10^{33}\,\mathrm{erg\,s^{-1}}$ from magnetically funneled accretion shocks \cite{Barrett1999}, while black widow pulsars have $L_X\sim10^{30}$--$10^{32}\,\mathrm{erg\,s^{-1}}$ from intrabinary shocks.
Neither \Jzero\ nor \Jone\ is detected by \textit{Swift}, with stringent upper limits of $L_X<7.1\times10^{29}\,\mathrm{erg\,s^{-1}}$ for \Jzero\ at 200 pc and $L_X<7.6\times10^{29}\,\mathrm{erg\,s^{-1}}$ for \Jone\ at 300 pc, which is below the expected luminosities of both polars and black widow systems.

\clearpage
%%%%%%%%%%%%%%%% EXTENDED DATA %%%%%%%%%%%%%%%

\section*{Extended Data}

\begin{figure}
    \centering
    \includegraphics[height=0.85\textheight]{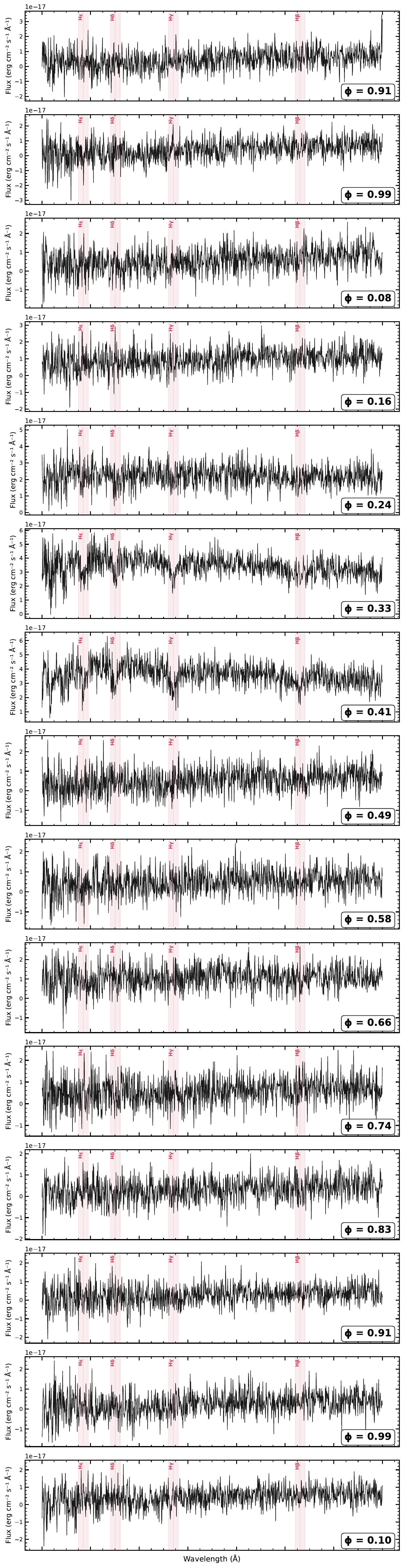}
    \caption{\textbf{Phase-resolved spectra of \Jone.} Transient Balmer absorption appears near peak brightness, consistent with absorption features from an accretion-powered hot spot rotating into view.}
    \label{fig:J1444keckall}
\end{figure}

\backmatter

\clearpage

\bmhead{Acknowledgements}
We explicitly acknowledge the contributions of our late co-author Tom Marsh for his early work on understanding these systems. We also acknowledge Tom Prince for his contributions to the Keck proposals used to collect the data presented in this work.
AH acknowledges support from the National Science Foundation Graduate Research Fellowship Program under Grant No. 2141064 and the MIT Dean of Science Fellowship. PR-G acknowledges support by the Spanish \textit{Agencia Estatal de Investigaci\'on} via PID2021--124879NB--I00 and PID2024--161863NB--I00.

\section*{Competing interests}
The authors declare no competing interests.

\section*{Data availability}
ZTF imaging data are available from the Zwicky Transient Facility public data release (\url{https://www.ztf.caltech.edu}). The \textit{SPHEREx} spectro-photometry and \textit{Swift}/XRT observations are available through their respective mission archives. The Keck/LRIS and Keck/ESI data are publicly available through the Keck Observatory Archive (\url{https://koa.ipac.caltech.edu}). The reduced HiPERCAM light curves and reduced spectra are publicly available at \url{https://github.com/aaronhouseholder/ZTFJ1444-and-ZTFJ0440-data}. Additional data products such as MCMC posterior samples are available from the corresponding author upon reasonable request.

\section*{Author contributions}
A.H. led the data analysis and modeling and wrote the manuscript. K.S. and K.B.B. first identified \Jone\ and \Jzero, and K.S. and K.B.B. led most of the follow-up observations of these objects. K.B.B. and T.R.M. initially developed the idea that such systems could exist based on the data for \Jzero\ and ZTF J1406+1222. S.A.R. helped A.H. develop the mass and radius constraints and stability criteria for these objects. K.E.-B. proposed for and collected the ESI data for \Jzero. All authors contributed to the scientific interpretation and to the preparation of the manuscript.

\section*{Correspondence}
Correspondence and requests for materials should be addressed to A.H. (aaron593@mit.edu).

\section*{Code availability}
The analysis was performed using a combination of custom Python code and publicly available software packages. The custom code will be archived in a public repository upon publication.

\bibliography{sn-bibliography}

\end{document}